\documentstyle[12pt]{article}
\begin{document}
\begin{titlepage}
  \title{Neutrino Oscillations Induced by Gravitational Recoil Effects}
\author{G. Lambiase$^{a,b}$\thanks{E-mail: lambiase@sa.infn.it} \\
{\em $^a$Dipartimento di Fisica E.R. Caianiello,} \\
 {\em Universit\'a di Salerno, I-84081 Baronissi (Sa), Italy.} \\
 {\em $^b$Istituto Nazionale di Fisica Nucleare, Sez. di Napoli, Italy.}
 \\ }
\date{\today}
\maketitle
\begin{abstract}
Quantum gravitational fluctuations of the space-time background,
described by virtual $D$ branes, may induce the neutrino
oscillations if a tiny violation of the Lorentz invariance (or a
violation of the equivalence principle) is imposed. In this
framework, the oscillation length of massless neutrinos turns out
to be proportional to $E^{-2}M$, where $E$ is the neutrino energy
and $M$ is the mass scale characterizing the topological
fluctuations in the vacuum. Such a functional dependence on the
energy is the same obtained in the framework of loop quantum
gravity.
\end{abstract}
\end{titlepage}

The attempts to build the quantum theory of gravity have clearly
shown that spacetime must have a non trivial topology at the Plank
scale. The suggestion that spacetime could have a foam-like
structure was, for the first time, advanced by Wheeler
\cite{wheeler}. In the last years, the study of quantum
fluctuations of the spacetime background has received a growing
interest and today it represents a very active research area in
physics \cite{hawking80,hawking82,ellis84,ellis92,garay,ellis99}.

In the presence of Plank-size topological fluctuations, the
quantum gravitational vacuum might behave as a non trivial medium.
Such a behaviour has been proven in the framework of string theory
\cite{amelino97} and of the canonical approach to quantum gravity
\cite{pullin} (see also \cite{yu}). In particular, we point out
the underlying idea of Ref. \cite{amelino97}: {\it quantum
gravitational fluctuations in the vacuum must be modified by the
passage of an energetic particle and the recoil will be reflected
in back reaction effects on the propagating particle} (see Ref.
\cite{ellis00}).

From an experimental setting, the present status eludes any
possibility to probe effects occurring at Plank energy.
Nevertheless, it has been recently suggested that $\gamma$-ray
bursts might be a possible candidate to test the theories of
quantum gravity \cite{amelino98}. The argumentations of the
authors are based on the peculiar physical properties of
$\gamma$-ray bursts, i.e. their origin at cosmological distance
and their high energy, which might make them sensitive to a
dispersion scale comparable with the Plank scale \cite{amelino98}.

In this note we investigate the possibility that the foamy
structure of the gravitational background, described by the
Ellis-Mavromatos-Nanopoulos-Volkov (EMNV) model \cite{ellis00a},
may induce the mixing of neutrinos if a tiny Lorentz invariance
violation is imposed \cite{coleman}. As a result, we find that the
inverse of the neutrino oscillation length does depend on the
square of the neutrino energy. This is the same functional
dependence derived in Ref. \cite{alfaro} by using the loop quantum
gravity approach (for a review, see \cite{rovelli}).

The EMNV model envisages virtual $D$ branes as responsible of the
foam structure of the space-time. The basic topic of this model is
the recoil effect of a $D$ brane struck by a boson
\cite{ellis99,ellis00} or a fermion \cite{ellis00a} particle,
which would induce an energy dependence of the background metric
through off-diagonal terms given by $G_{0i}\sim u_i$. Here $u_i$
is the average recoil velocity of the generic $D$ brane
\cite{ellis00a}, and it is of the order $u_i\sim E/M\ll 1$, where
$E$ is the energy of the particle scattering off the $D$ brane,
and $M$ has the dimension of a mass which characterizes the
quantum fluctuations scale. As a consequence of the off-diagonal
term in the metric tensor ($G_{0i}$), the Lorentz invariance is
broken.

Without pretending to be exhaustive, we just recall the main
features of the ENMV model necessary for our considerations (for
details, see Ref. \cite{ellis00a} and references therein).

The momentum conservation during the recoil process implies a
energy dependence of the metric tensor whose asymptotic (in the
time) components are (for a $D$-dimensional spacetime)
\begin{equation}\label{metric}
  G_{ij}=\delta_{ij}\,,\quad G_{00}=-1\,,
\end{equation}
 \[
  G_{0i}\sim u_i\,, \quad i, j=1, \ldots , D-1\,,
 \]
resulting constant in the spacetime coordinates. The metric
(\ref{metric}) induces a variation of the light velocity given by
$\delta c/c\sim -E/M$.

In view of the application to neutrino oscillations, we are
interested to the recoil effects on fermions. A massless fermion
is described by the covariant Dirac equation in curved spacetime
 \begin{equation}\label{dirac0}
 [i\gamma^{\mu}(\nabla_{\mu}-\Gamma_{\mu})]\psi=0\,.
 \end{equation}
The general relativity matrices $\gamma^{\mu}$ are related to the
Lorentz matrices $\gamma^m$ by means of the vierbein fields
$e_{\mu}^m$ ($\gamma_\mu=e_\mu^m\gamma_m$), with components given
by
 \[
 e^\mu_m=\left(\matrix{ -1 & 0 & 0 & -u_1 \cr
                                0 & -1 & 0 & -u_2 \cr
                                0 & 0 & -1 & -u_3 \cr
                                  0 & 0& 0 & 1 \cr }\right)\,.
 \]
The operator $\nabla_{\mu}$ is the usual covariant derivative,
$\nabla_{\mu}=\partial_\mu+\Gamma^\nu_{\mu\nu}$, and
$\Gamma_{\mu}$ are the spin connections defined as
 \[
 \Gamma_{\mu}={1\over
 8}[\gamma^m, \gamma^n]e^{\nu}_me_{\nu n;\mu}
 \]
(semicolon represents the covariant derivative). In terms of the
metric (\ref{metric}), the Dirac equation (\ref{dirac0}) becomes
\begin{equation}\label{dirac}
  [\gamma^m\partial_m-\gamma^0(u_i\nabla_i)]\psi=0\,.
\end{equation}
The action of the operator $\gamma^\mu\partial_\mu$ allows to
derive the dispersion relation \cite{ellis00a}
\begin{equation}\label{dispersion}
  E^2=p^2-2(u_ip_i)E\,,
\end{equation}
where, as already noted, $\vert {\vec u}\vert\sim E/M\ll 1$ and
$p$ is the momentum of the massless fermion. As argued in
\cite{ellis00a}, if $\gamma$-rays bursts may emit pulses of
neutrinos with $E\sim 10^{10}$GeV, the effects of the second term
in (\ref{dispersion}) could be tested provided $M\sim 10^{27}$GeV.

Let us now assume that the $u$-term in (\ref{dispersion}) violates
the Lorentz invariance. As argued by Coleman and Glashow
\cite{coleman} (see also {\cite{coleman1}), this would imply that
different species of neutrino massless may have different maximal
attainable speeds that does not coincide with the light velocity
$c$. In such a case, neutrino oscillations can occur if the
neutrino flavor eigenstates $\nu_f$, $f=e, \mu$, are linear
superposition of the velocity eigenstates $\nu_a$, $a=1, 2$, at
infinite momentum. The relation between velocity and flavor
eigenstates is
  \begin{eqnarray}
 \nu_\mu &=& \cos\theta \,\nu_1+\sin\theta\,\nu_2\,, \nonumber \\
  \nu_e &=& \cos\theta \,\nu_2-\sin\theta\,\nu_1\,, \nonumber
 \end{eqnarray}
where $\theta$ is the mixing angle.

{} From Eq. (\ref{dispersion}) it follows that the velocity $v_a$
of the neutrinos eigenstates $\nu_a$ can be written as (we
consider one-dimensional motion)
\begin{equation}\label{velocity}
  v_a\sim 1+u_a= 1+\alpha_a\,\frac{E}{M}\,,\quad a=1,2\,,
\end{equation}
where the notation $u_a=\alpha_a E/M$ has been introduced. The
parameter $\alpha_a$ characterizes the maximal attainable velocity
of neutrinos $\nu_a$.

During their evolution, neutrinos propagate as a linear
combination of velocity eigenstates whose energies are $E_a$,
$a=1, 2$. Taking into account Eq. (\ref{velocity}), one infers
that energy difference is \cite{coleman}
\begin{equation}\label{energydifference}
  E_1-E_2=\delta v E\sim \delta \alpha \,\frac{E^2}{M}\,.
\end{equation}
In Eq. (\ref{energydifference}), $\delta v=v_1-v_2$ and $\delta
\alpha=\alpha_1-\alpha_2$. Thus, $\delta \alpha$ {\it measures}
the degree of violation of the Lorentz invariance.

The probability that the neutrino preserves its flavor is
therefore
\begin{equation}\label{probability}
  P=\sin^2\theta\sin^2\left(\delta\alpha \,\frac{ E^2}{2M}\, L\right)\,,
\end{equation}
being $L$ the cosmological distance traveled by neutrinos between
the emission and detection. From (\ref{probability}) one infers
the inverse of the oscillation length
\begin{equation}\label{osclength}
  \lambda^{-1}_{osc}\sim \delta \alpha\, \frac{E^2}{M}\,.
\end{equation}
Bounds on $\delta \alpha$ can be estimated by using the relation
$\lambda_{osc}\leq L$. For example, the values for the energy
$E\sim 10^{10}$GeV and the mass scale $M\sim 10^{27}$GeV, the same
of Ref. \cite{ellis00a}, and $L\sim 2.7\times 10^{22}$km yield
$\delta \alpha \geq 10^{-34}$.

The energy dependence of the oscillation length of neutrinos, Eq.
(\ref{osclength}), is the same of Ref. \cite{alfaro}. There the
propagation of fermion fields (neutrinos) has been studied in the
framework of the loop quantum gravity \cite{rovelli}. In that
case, the oscillation length of neutrinos is given by
$\lambda_{osc}^{-1}=\Delta \rho_1 E^2/M_P$, where the mass scale
is characterized by the Plank mass $M_P$, and $\Delta \rho_1$
measures a violation of the equivalence principle \cite{alfaro}.

It is worth noting that the possibility to generate neutrino
oscillations (even for massless neutrinos) via the violation of
the equivalence principle as in Refs. \cite{GAS,HAL}, could also
apply in the EMNV model. Results are the same discussed in this
note (one obtains the oscillation length given by Eq.
(\ref{osclength})) and do coincide with ones of Ref.
\cite{alfaro}.

\vspace{0.5in}

\centerline{Acknowledgments}

Research supported by fund MURST PRIN 99.

\end{document}